# Geostationary Earth Orbit Satellite Model using Easy Java Simulation

Loo Kang WEE[1], Giam Hwee GOH[2]

[1]Ministry of Education, Education Technology Division, Singapore
[2]Ministry of Education, Yishun Junior College, Singapore

wee_loo_kang@moe.gov.sg, goh_giam_hwee@moe.edu.sg

Abstract: We develop an Easy Java Simulation (EJS) model for students to visualize geostationary orbits near Earth, modeled using Java 3D implementation of the EJS 3D library. The simulated physics model is described and simulated using simple constant angular velocity equation. Four computer model design ideas such as 1) simple and realistic 3D view and associated learning to real world, 2) comparative visualization of permanent geostationary satellite 3) examples of non-geostationary orbits of different 3-1) rotation sense, 3-2) periods, 3-3) planes and 4) incorrect physics model for conceptual discourse are discussed. General feedback from the students has been relatively positive, and we hope teachers will find the computer model useful in their own classes.

## I. INTRODUCTION

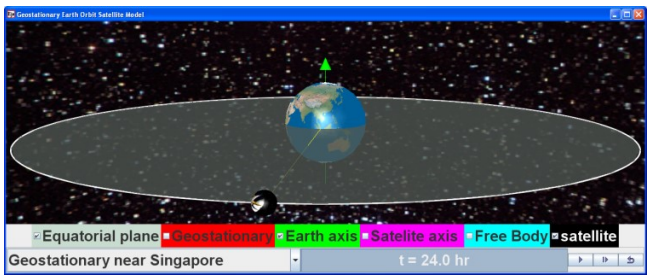

Figure 1. EJS applet view of the simulation learning environment showing the orbital view of Earth and a geostationary satellite in Java 3D implementation of the EJS 3D library with a bottom control panel for inquiry activities.

Geostationary orbits are inaccessible physics concepts that are difficult to experience in real life as the students need to be in outer space near the Earth, which we argue is near impossible financially, so as to observe the motion of satellites relative to Earth.

In traditional classrooms, students could be asked to imagine Earth and the geostationary satellites, or watch YouTube videos of pre-defined geostationary orbit with little or no opportunity for students to explore and understand cases when the orbits are non-geostationary. Thus we feel there is justification to use computer model to support active experiential [1] student-centered learning.

We created a computer model, also known as simulation, to allow our students to visualize the phenomena, using a free authoring toolkit called Easy Java Simulation (EJS) [2]. This simulation utilized a new feature in EJS using new Java 3D implementation as unveiled during Multimedia in Physics Teaching and Learning Conference MPTL 14 [3]. This new Java 3D implementation allows ordinary teachers to create realistic 3D tools for physics education.

In addition, our quick literature review suggests we are probably a world first in Physics Education journals to come up with a simulation tool to help students learn geostationary orbits that is customized to GCE Advanced Level Physics [4].

Building on open source codes shared by the Open Source Physics (OSP) community like, Francisco's "Examples Earth and Moon 3 D" [5], and with help from Fu-Kwun's NTNUJAVA Virtual Physics Laboratory [6], we customized an Easy Java Simulation (EJS) computer model as a 3D visualization tool (Figure 1), downloadable from https://sites.google.com/site/lookang/edulabgravityearthandsatelliteyjc/ejs_EarthAndSatelite.jar?attredirects=0&d=1 digital libraries in ComPadre Open Source Physics [7] and NTNUJAVA Virtual Physics Laboratory [8], creative commons attribution licensed.

The recommended system requirement for running this EJS model in Java 3D is the Intel Pentium processor.

## II. Physics model

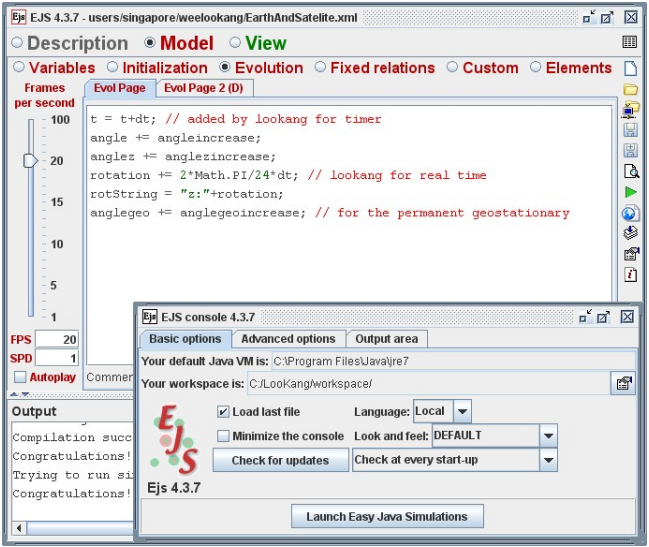

Figure 2. EJS authoring tool view at the 'Evolution Page' showing equations (1) and (2) with the Evolution Page as ordinary Euler solver statements.

In this model as implemented in EJS (Figure 2), the rotating Earth as well as the geostationary satellite is governed by a simple constant angular velocity equation (1) about the $z$ axis, where $\vartheta$ is the 'angle' of rotation of Earth and the constant on the right hand side of the equation (1) assume time in hours.

$$\frac{d\vartheta}{dt} = \frac{2\pi}{24} \tag{1}$$

Users who wish to model non-geostationary orbits of different angular velocity $\vartheta_k$ need to modify equation (1) into equation (2) with

$k = -1.0$ for opposite rotational direction of geostationary satellite

$k = 0.5$ for half the geostationary angular velocity

$k = 2.0$ for double the geostationary angular velocity.

$$\frac{d\vartheta_k}{dt} = \frac{2\pi\,k}{24} \tag{2}$$

Advanced users who wish to model other non-geostationary orbits involving the rotation about $x$ or $y$ axes need to introduce another angle say 'angle z' using equation (1).

Thus, the satellite will need to be drawn with $x$, $y$ and $z$ coordinates using equation (3) in relations to the angle $\vartheta$, angle z, $\vartheta_z$ and $R$ the geostationary distance from centre of Earth to the satellite.

$$\begin{pmatrix} x \\ y \\ z \end{pmatrix} = \begin{pmatrix} R\cos\vartheta\cos\vartheta_z \\ R\sin\vartheta\cos\vartheta_z \\ R\sin\vartheta_z \end{pmatrix} \tag{3}$$

This physics model when implemented in a computer model allows users to explore the simulation productively [9] serving as a powerful visualization tool for learning.

Beyond the scope of this paper, expert users who wish to model gravitational equations should refer to other models [10, 11] that use Newton's gravitational force and initial velocities to predict satellites' motions.

### III. Four computer model Design ideas

To add to the body of knowledge surrounding why simulations could be effective tools, we share four computer model design ideas-insights that we believed have raised the effectiveness and usefulness of the tool for students' active learning.

#### A. Simple and realistic 3D view and associated learning to real world

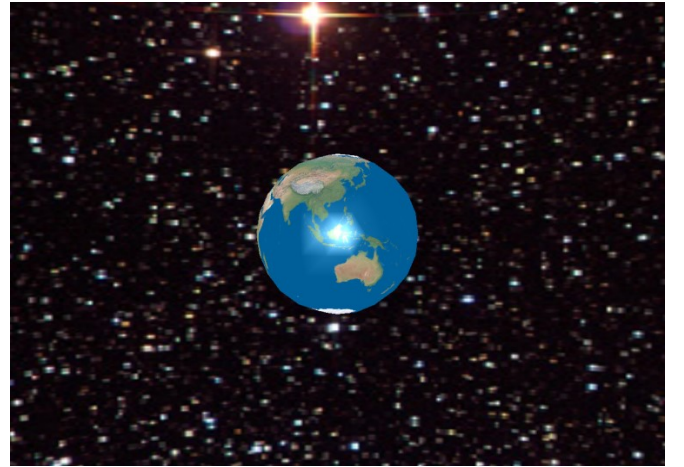

Figure 3. Orbital view of Earth as a 3D sphere and the universe in the background of the 3D drawing panel in EJS implemented in Java 3D.

To achieve a realistic view of Earth, we used a texture map that can be found on the public domain of the internet such as the Natural Earth III by Tom [12]. We further used a free graphics software GIMP to reduce the size of the canvas dimensions to 2000x1000 pixels arbitrarily to achieve a reasonable file size of 247 kB to be used in EJS 3D sphere object as texture.

To create a view of the universe viewed from Earth, we use a texture map that can be found on NASA website [13] inserted into the 3D drawing panel with graphical aspect referring to that file.

In the 3D drawing panel in EJS, select the JAVA 3D implementation instead of the simple 3D to produce a view of Earth and the universe (Figure 3).

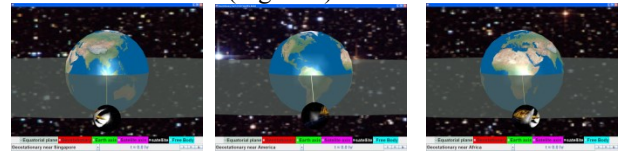

Figure 4. Orbital view of Earth with geostationary satellite above South-East Asia (Singapore), Americas and Africa continents for associated learning to real world.

To create associated learning to real world reference position, we create three positions with land mass around the equator such as South East Asia (Singapore), Africa and Americas continent. We use equation (4) to position these three points of geostationary orbits where $\vartheta$ is the 'angle' where

$\vartheta_0 = \vartheta_{Singapore} \approx 0.25$ , $\vartheta_{America} \approx 3.6$ and $\vartheta_{Africa} \approx 5.2$ radians respectively (Figure 4) and move in an angular velocity equal to the Earth's rotation $\frac{2\pi}{24}$ in hours.

$$\vartheta_{k=1} = \vartheta_0 + \frac{2\pi}{24} dt \qquad (4)$$

To aid in the visualization from different perspectives, a semi transparent equatorial plane and axis of rotations for the Earth and satellite were added in to help students gain a better visualization of the 3D space that they are interacting with.

To further add realism, the radius of Earth, $R_{Earth}$ and the orbits $R$, were modeled by drawing on data collected from validated sources like NASA website and Wikipedia. We used a scale of $1 \times 10^6$ m to represent the radius of Earth in the model, $R_{Earth} \approx 0.637$ and radial distance from centre of Earth to geostationary orbit, $R \approx 4.23$ .

Our internet scan found some simulations, represented in only 2D [14] or commercial 3D [15], which show specifically geostationary orbits but are more complicated to use. Thus, we believe our computer model could be a simpler tool for teachers and students to use for visualization of geostationary orbits around Earth.

### B. Comparative visualization of permanent geostationary satellite

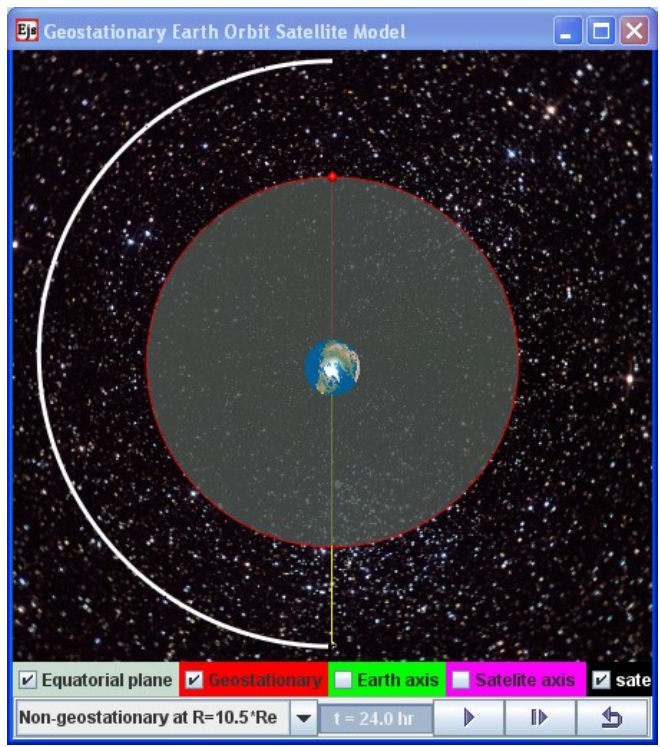

Figure 5. Orbital view of Earth from the North Pole with (**red**) geostationary satellite above South-East Asia (Singapore) completing one revolution after 24 hours, while a non-geostationary orbit at radius $R = 10.5*R_{Earth}$ has completed half a revolution.

We found that the provision of a timer '$t$ = XX hrs' alone was insufficient for students to make a clear meaning of non-geostationary orbits through observation. The act of observing the numerical value in the timer, say 48 hours (Figure 5) for an orbit that is twice the period of the rotation of the Earth can be better enhanced by showing another geostationary object simultaneously, in our case for example, a red satellite above Singapore.

Thus, a checkbox was designed for students to activate the display of a permanent example of a geostationary orbit so that quick and clear interpretation of the meaning of non-geostationary by comparing with a geostationary orbit could be made.

### C. Examples of non-geostationary orbits

To further improve understanding of the geostationary orbits, we also designed typical non-geostationary orbits for students to compare the differences.

#### 1) Non-geostationary due to direction (rotation sense)

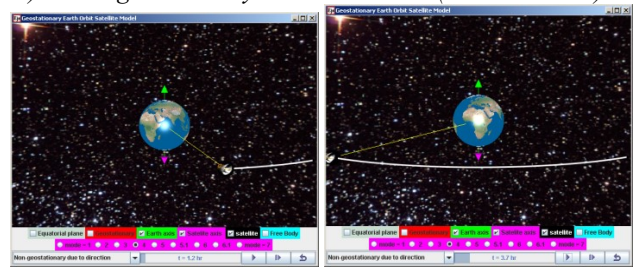

Figure 6. Orbital view of Earth with an (**white**) ecliptical orbit path showing a satellite rotating in the sense of (**magenta**) axis of rotation while the Earth rotate in the sense of the (**green**) axis. The left to right image illustrate the time lapse visualization where the point below the satellite on Earth is always changing, thus not geostationary.

Some students do not appreciate that for geostationary orbit to occur, it not only requires a period of 24 hours and lies on the equator plane of the Earth, it also need to be rotating in the same direction as the Earth's rotation about its own axis (Figure 6). The comparative visualization in the rotating Earth and orbiting satellite, as well as the 2 axis of rotations served to highlight the importance of the geostationary orbit's rotation sense.

#### 2) Non-geostationary circular motion at different periods

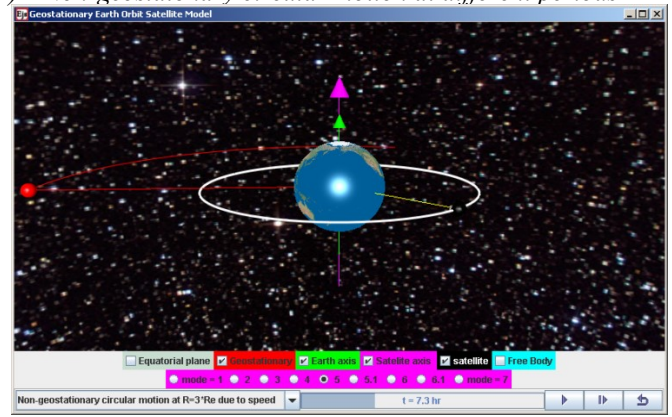

Figure 7. Orbital view of Earth non-geostationary cirular orbit of $R = 3*R_{Earth}$ with a period of 7.3 hours, in the equator plane and same rotation sense as Earth.

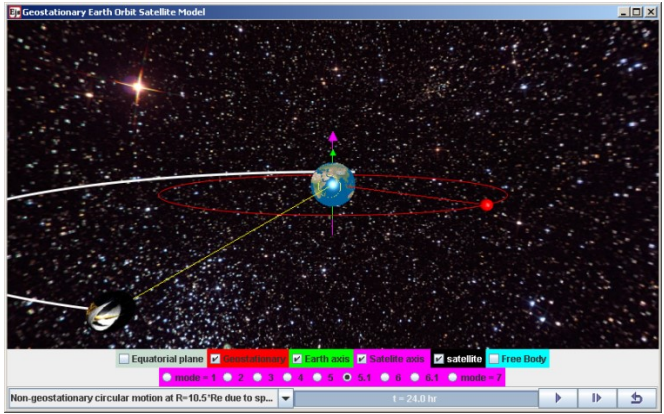

Figure 8. Orbital view of Earth non-geostationary cirular orbit of R = 10.5*$R_{Earth}$ with a period of 48 hours, in the equator plane and same rotation sense as Earth.

To illustrate that the period's of geostationary orbit needs to be 24 hours or its angular speed needs to be equal to that of Earth's, 2 non-geostationary orbits were designed on the equator plane and with the same rotation sense as Earth's, but one orbiting faster at an orbital radius three times that of Earth's radius (Figure 7) and the other orbiting slower at an orbital radius 10.5 times that of Earth's radius (Figure 8).

One teaching point here was to introduce the free body diagram of the satellite and equating the net force on the satellite to its centripetal force (equation 5 and 6 respectively) to derive the predicted value of the period of $T_{R=3Rearth}$ = 7.3 hours and $T_{R=10.5Rearth}$ = 48.0 hours respectively. Here students need to know that univeral gravitational constant $G$ = 6.67 × $10^{-11}$ m³kg⁻¹s⁻², mass of Earth $M_{Earth}$ = 5.97 × $10^{24}$ kg and the mass of satellite, $m$ were independent in the determination of the period $T$. In addtion, the conversion of seconds to hours proved to be challenging to novice students as well.

$$G\frac{mM_{Earth}}{3R_{Earth}} = m(3R_{Earth})(\frac{2\pi}{T_{R=3R_{Earth}}})^2 \tag{5}$$

$$G\frac{mM_{Earth}}{10.5R_{Earth}} = m(10.5R_{Earth})(\frac{2\pi}{T_{R=10.5R_{Earth}}})^2 \tag{6}$$

### 3) Non-geostationary due to plane such as polar orbits with period = 24 hours

Now that we have covered the need of same rotation sense in 1) and same period as Earth or angular speed in 2), lastly a polar orbit serves as an example for need to be on Earth's equator plane (Figure 9).

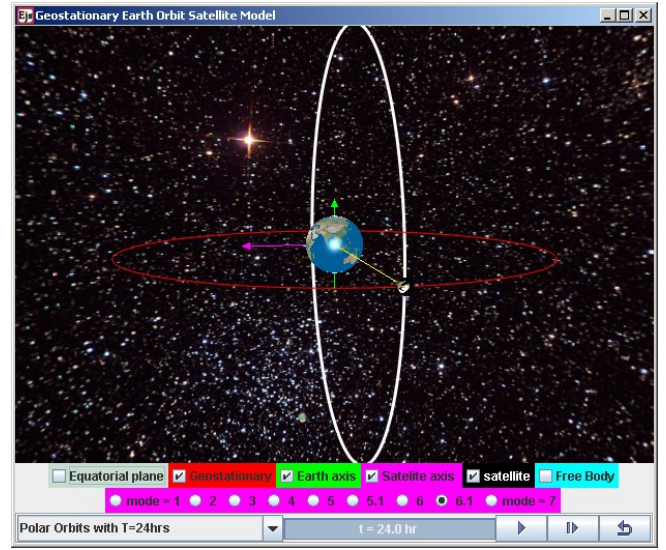

Figure 9. Orbital view of Earth non-geostationary polar orbit with period T = 24 hours, in a plane containing the north and south pole (white trail) different from the equatorial plane geostationary oribt (red trail).

Here, we typically shared with students the real life applications of polar orbit such as for earth mapping or observation capturing the Earth as time passes as well as some weather satellites, to make the physics knowledge relevant to daily life.

### D. Incorrect physics for conceptual reasoning

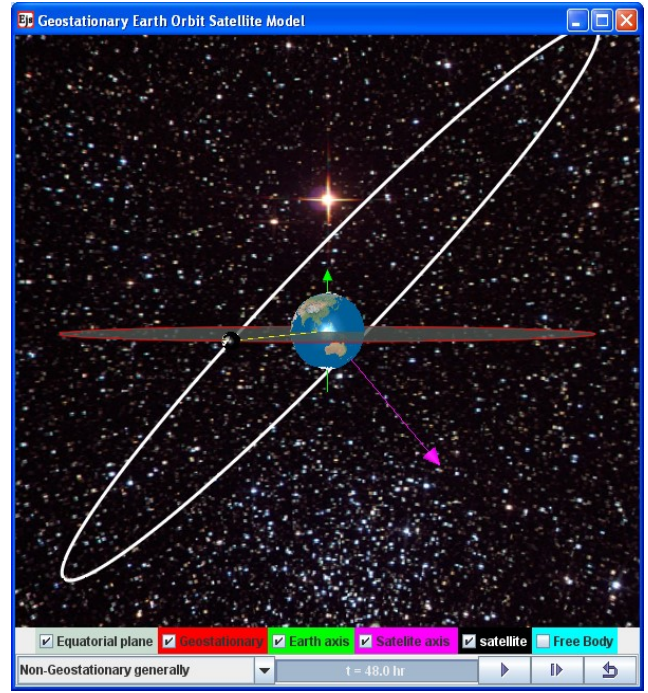

Figure 10. Orbital view of Earth with an (**white**) ecliptical orbit path with its (**magenta**) axis of rotation while showing the geostatioanry orbit and its (**green**) axis of rotation.

To represent a more general orbit not in the plane of the equator, we use equation (3) as mention in II. This produced a simplistic orbit that can be used to allow visualization of non-geostationary orbits that is not in the plane of the equator of Earth's rotation (Figure 10). By having incorrect physics [16] motion represented, we challenged students to explain and elaborate what is incorrect about this orbit, for instance in relation to its higher speed when nearer to Earth and lower speed when further away.

Lastly, an unlikely orbit (Figure 11) that appears to be geostationary above a point on the northern hemisphere of Earth was used to challenge thinking about what is 'wrong' with this orbit. A free body diagram showing the equal and opposite forces acting separately on the Earth and satellite helped students to use what they learnt about Newton's Third Law in this context. Students explained that the direction of the force was towards the Earth's center while in a circular orbit, would require that the net force to be pointed towards the center of the circular path, which was not happening in this mode. We also got the students to think about the force that needs to be continually applied on the satellite in order for this unlikely orbit to be possible.

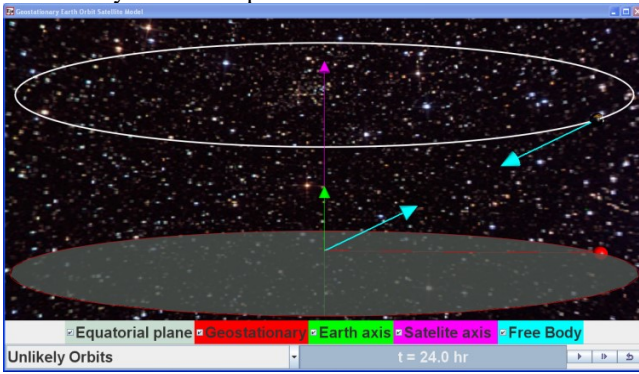

Figure 11. A free body diagram of Earth and satelite showing the forces (teal) on the Earth and satellite as equal and opposite in direction acting on different bodies.

## IV. FEEDBACK FROM STUDENTS

We include excerpts from the qualitative survey results and informal interviews with the students to give some themes and insights into the conditions and processes during the laboratory lessons. Words in brackets < > are added to improve the readability of the qualitative interviews.

### 1) Improved 3D visualization from different perspectives

"The lesson allows me to understand the movement of a satellite which we cannot see normally in real life and are unable to comprehend from the 2D <textbook> diagram. Thus, the 3D simulation allows me to learn better".

"Allow <me> to get a better understanding of the topic as simulation aids in visualizing the various questions easily, thus, able to solve the question. The lessons give me a clearer explanation of how things works thus, allowing me to understand".

### 2) Need for strong inquiry learning activities

"Not enough group activities. The questions are not interesting enough".

"Not focusing on specific questions, and just focusing on concepts".

We suspect that the students are requesting for more open-ended inquiry activities to be designed for learning instead of embedding the simulation to existing tutorial questions that tend to be more theoretical and only test subset of the knowledge regarding geostationary orbits.

### 3) Need well designed simulation [9]

Some students suggest having "more diverse options" so that the simulations could be more interesting and appealing.

This suggestion has inspired us to expand on the option of non-geostationary orbits due to speed by including a higher angular speed at $3*R_E$ and a lower angular speed at orbit radius of $10.5*R_E$.

### 4) Appreciative learners

"I would like to show my appreciation for the <information and communication technology> ICT inventors and teachers who participated in this ICT learning programme as it is a new opportunity for us to pick up high technology skills to pick up physics."

"I would like to thank my teacher for allowing us to gain exposure to these simulations and how they are able to be used to allow us <to> understand the topic better."

## V. CONCLUSION

The simplified constant angular velocity physics model is discussed and implemented in EJS and the equations (1) to (3) give a brief account of how to create a geostationary orbit simulation tool. Despite using only simple constant velocity equation, we were able to create this tool for the learning of geostationary orbits for physics education. The full computer model can be downloaded from https://sites.google.com/site/lookang/edulabgravityearthandsatelliteyjc/ejs_EarthAndSatelite.jar?attredirects=0&d=1 ComPadre Open Source Physics [7] and NTNU Virtual Physics Laboratory [8] digital libraries.

Four computer model design ideas such as 1) simple and realistic 3D view and associated learning to real world, 2) comparative visualization of a permanent geostationary satellite 3) examples of non-geostationary orbits of different 3-1) rotation sense, 3-2) periods and 3-3) plane and 4) intentionally incorrect physics model [16] for conceptual discourse. We implemented these design-ideas in our EJS model that we believe can further support student learning.

From the learning theory of constructionism, teachers could get interested students to go through the process of model building as a pedagogical tool [17], for instance as a project assignment, heightens the relevance of this article's model construction details.

General feedback from the students has been relatively positive, triangulated from the survey questions, interviews with students and discussions with teachers and we hope more teachers will find the simulation useful in their own classes.

## ACKNOWLEDGEMENT

We wish to acknowledge the passionate contributions of Francisco Esquembre, Fu-Kwun Hwang and Wolfgang Christian for their ideas and insights in the co-creation of interactive simulation and curriculum materials.

This research is made possible; thanks to the eduLab project NRF2011-EDU001-EL001 Java Simulation Design for Teaching and Learning, awarded by the National Research

Foundation, Singapore in collaboration with National Institute of Education, Singapore and the Ministry of Education (MOE), Singapore.

Lastly, we also thank MOE for the recognition of our research on the computer model lessons as significant innovation in 2012 MOE Innergy (HQ) GOLD Awards by Educational Technology Division and Academy of Singapore Teachers.

AUTHOR

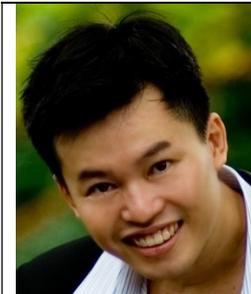

Loo Kang WEE is currently an educational technology specialist at the Ministry of Education, Singapore. He was a junior college physics lecturer and his research interest is in Open Source Physics tools like Easy Java Simulation for designing computer models and use of Tracker.

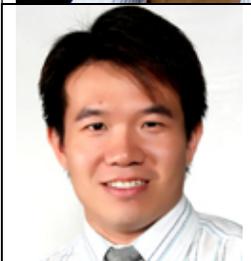

Giam Hwee GOH is currently the Head of Science Department in Yishun Junior College, Singapore. He teaches Physics to both year 1 and 2 students at the college and advocates inquiry-based science teaching and learning through effective and efficient means.